\documentclass[%
 preprint,
superscriptaddress,
 amsmath,amssymb,
 aps,
]{revtex4-2}

\usepackage{graphicx}
\usepackage{amssymb}
\usepackage{dcolumn}
\usepackage{bm}

\usepackage{changes}
\definechangesauthor[name={Oliver Gessner}, color=orange]{OG}
\definechangesauthor[name={Wolfgang Eberhardt}, color=blue]{WE}
\definechangesauthor[name={Stefan Duester}, color=red]{SD}
\definechangesauthor[name={Serguei Molodtsov}, color=green]{SM}
\definechangesauthor[name={Nikolay Kabachnik}, color=magenta]{NK}
\definechangesauthor[name={Friedrich Roth}, color=brown]{FR}
\definechangesauthor[name={Lukas Wenthaus}, color=teal]{LW}

\usepackage{subcaption}

\begin{document}

\title{New insights into the laser-assisted photoelectric effect from solid-state surfaces\\
}

\author{Lukas Wenthaus}
\email{lukas.wenthaus@desy.de}
\affiliation{Deutsches Elektronen-Synchrotron DESY, Notkestrasse 85, 22603 Hamburg, Germany}%

\author{Nikolay M. Kabachnik}
\affiliation{European XFEL GmbH, Holzkoppel 4, 22869, Schenefeld, Germany}%
\affiliation{Donostia International Physics Center(DIPC), E-20018 San Sebastian/Donostia, Spain}%

\author{Mario Borgwardt}
\affiliation{Chemical Sciences Division, Lawrence Berkeley National Laboratory, Berkeley, California 94720, USA}%
  
\author{Steffen Palutke}
\affiliation{Deutsches Elektronen-Synchrotron DESY, Notkestrasse 85, 22603 Hamburg, Germany}%

\author{Dmytro Kutnyakhov}
\affiliation{Deutsches Elektronen-Synchrotron DESY, Notkestrasse 85, 22603 Hamburg, Germany}%

\author{Federico Pressacco}
\affiliation{Deutsches Elektronen-Synchrotron DESY, Notkestrasse 85, 22603 Hamburg, Germany}%

\author{Markus Scholz}
\affiliation{Deutsches Elektronen-Synchrotron DESY, Notkestrasse 85, 22603 Hamburg, Germany}%

\author{Dmitrii Potorochin}
\affiliation{Institute of Experimental Physics, TU Bergakademie Freiberg, Leipziger Strasse 23, 09599 Freiberg, Germany}%
\affiliation{Center for Efficient High Temperature Processes and Materials Conversion (ZeHS), TU Bergakademie Freiberg, Winklerstrasse 5, 09599, Freiberg, Germany}%

\author{Nils Wind}
\affiliation{Universit\"at Hamburg, Luruper Chaussee 149, 22761, Hamburg, Germany}%

\author{Stefan D\"usterer}
\affiliation{Deutsches Elektronen-Synchrotron DESY, Notkestrasse 85, 22603 Hamburg, Germany}%

\author{G\"unter Brenner}
\affiliation{Deutsches Elektronen-Synchrotron DESY, Notkestrasse 85, 22603 Hamburg, Germany}%

\author{Oliver Gessner}
\affiliation{Chemical Sciences Division, Lawrence Berkeley National Laboratory, Berkeley, California 94720, USA}%

\author{Serguei Molodtsov}
\affiliation{European XFEL GmbH, Holzkoppel 4, 22869, Schenefeld, Germany}%
\affiliation{Institute of Experimental Physics, TU Bergakademie Freiberg, Leipziger Strasse 23, 09599 Freiberg, Germany}%
\affiliation{Center for Efficient High Temperature Processes and Materials Conversion (ZeHS), TU Bergakademie Freiberg, Winklerstrasse 5, 09599, Freiberg, Germany}%

\author{Wolfgang Eberhardt}
\affiliation{Deutsches Elektronen-Synchrotron DESY, Notkestrasse 85, 22603 Hamburg, Germany}%

\author{Friedrich Roth}
    \email{friedrich.roth@physik.tu-freiberg.de}
\affiliation{Institute of Experimental Physics, TU Bergakademie Freiberg, Leipziger Strasse 23, 09599 Freiberg, Germany}%
\affiliation{Center for Efficient High Temperature Processes and Materials Conversion (ZeHS), TU Bergakademie Freiberg, Winklerstrasse 5, 09599, Freiberg, Germany}%

\date{\today}

\begin{abstract}

Photoemission from a solid surface provides a wealth of information about the electronic structure of the surface and its dynamic evolution. Ultrafast pump-probe experiments are particularly useful to study the dynamic interactions of photons with surfaces as well as the ensuing electron dynamics induced by these interactions. Time-resolved laser-assisted photoemission (tr-LAPE) from surfaces is a novel technique to gain deeper understanding of the fundamentals underlying the photoemission process. Here, we present the results of a femtosecond time-resolved soft X-ray photoelectron spectroscopy experiment on two different metal surfaces conducted at the X-ray Free-Electron Laser FLASH in Hamburg. We study photoemission from the W 4f and Pt 4f core levels using ultrashort soft X-ray pulses in combination with synchronized infrared (IR) laser pulses. When both pulses overlap in time and space, laser-assisted photoemission results in the formation of a series of sidebands that reflect the dynamics of the laser-surface interaction. We demonstrate a qualitatively new level of sideband generation up to the sixth order and a surprising material dependence of the number of sidebands that has so far not been predicted by theory. We provide a semi-quantitative explanation of this phenomenon based on the different dynamic dielectric responses of the two materials. Our results advance the understanding of the LAPE process and reveal new details of the IR field present in the surface region, which is determined by the dynamic interplay between the IR laser field and the dielectric response of the metal surfaces. 

\end{abstract}

\maketitle

\section{\label{sec:Introduction}Introduction}

Photoemission of electrons from atoms, molecules, and condensed matter provides the experimental basis of our understanding of electronic structure. Over the last fifty years, photoelectron spectroscopy (PES) has increasingly become one of the most widely used methods for materials characterization and investigation, and is now an essential analytical tool for probing the properties and working principles of materials and their functional interfaces. The technique combines high spectral resolution with the momentum selectivity and atomic-site specificity of valence- and core-electron emission. It can thus provide direct information on electronic band dispersions in energy-momentum space as well as on the local chemical and structural environment of atoms.

Over the past $\sim\,$20 years, time-domain X-ray spectroscopy techniques have gained increasing attention in a variety of scientific communities due to their potential to reveal details of fast dynamic processes in matter that are not accessible in static X-ray experiments or by optical time domain probes. By using the pulsed time structure of X-ray light sources, pump-probe schemes have been introduced to probe local electronic structure changes.

Among the various available X-ray probe schemes, time-resolved photoelectron spectroscopy (tr-XPS) is particularly well suited to study ultrafast processes in simple atoms and molecules \cite{Calegari_2014} as well as charge transfer dynamics in complex solid-state systems \cite{roth_2021, borgwardt_2020}. In tr-XPS, a pump pulse typically excites a ground state configuration of the system under investigation, which is monitored by a time-delayed X-ray probe pulse, mapping the systems internal electronic dynamics onto an energy-resolving detector.

When an electron is photo-emitted from an atom, molecule or solid in the presence of an intense optical laser field, an additional interaction can be observed. Whenever the X-ray pulse and the laser pulse coincide in time and space, the presence of the optical laser field modifies the photoelectron spectrum, leading to a redistribution of photoemission signal intensity and to the generation of multiple, equidistant peak structures termed sidebands (SBs). These replicas of the main photoemission peak are energetically separated from the main line by integer multiples of the laser photon energy. This two-color photoionization phenomenon is called the laser-assisted photoelectric effect (LAPE) and can be described as the simultaneous absorption or emission of one or more optical laser photons along with the X-ray pulse.

Although sidebands in solids have previously been used to determine the temporal overlap of X-ray and IR lasers in time-resolved photoemission experiments, a precise description and fundamental understanding of the intricate details of the involved processes is still missing. This is of crucial importance for many FEL-based experiments, since the appearance of sidebands is often part of the dynamic change in the recorded spectra that needs to be taken into account. Moreover, dynamic screening processes at metal surfaces are known to modify the microscopic local fields of the materials under investigation.

The redistribution of photoelectrons into sidebands can be seen as a spectral distortion overlapping with the actual physical effect of interest such as ultrafast photopeak shifts induced by charge transfer processes \cite{borgwardt_2020,arion_2015,roth_2019,neppl_2021}. The coincidence of non-resonant and resonant effects requires special attention when modeling the underlying physics to disentangle the different contributions. This is particularly important as emerging light source technologies provide ever more brilliant and shorter X-ray pulses that will, with simultaneous improvements on the detector side, lead to considerably higher spectral and temporal resolution experiments. Furthermore, the need for correspondingly shorter optical laser pulses matching the X-ray pulse properties will increase peak intensities so that, even at moderate photon fluences, laser-induced shifts (e.\,g., AC Stark shifts) and pronounced LAPE effects become apparent.

Most reported LAPE measurements on solid and liquid systems were conducted using high-harmonic generation (HHG)  light sources with comparatively low photon energies. These typically limit the spectral window to valence excitations that are often challenging to interpret due to their spectrally broad nature and complex structure \cite{miaja-avila_2006,saathoff_2008}. Combined with the relatively low kinetic energy of the photoelectrons, a precise modeling of the LAPE effect so far remained elusive. Specifically, the broadening of the sideband structures and the asymmetry of sideband intensities described by the LAPE response function (RF) \cite{miaja-avila_2006} could not be experimentally validated.\\

This work presents the first systematic investigation of dynamic screening by time-resolved, laser-assisted XPS of shallow core levels of metallic solids, which is employed to monitor  local electric fields at clean Pt\,(111) and W\,(110) single crystal surfaces. The experiment was conducted at the X-ray Free-Electron Laser FLASH in Hamburg \cite{ackermann_2007} using a photon energy of 514.5\,eV. Comparison of the sideband formation in these two metals measured under identical experimental conditions reveals an unexpected dependence of the number of sidebands on the target material. The significantly different responses are attributed to electronic screening and the modification of the local electric fields at the clean Pt\,(111) and W\,(110) single crystal surfaces.
The simple spectral shape and the high kinetic energy of core-level photoelectrons allow us to gain a deeper understanding and improve the modeling of the LAPE effect to derive a semi-quantitative explanation of the observed phenomenon. Moreover, the observed sensitivity of tr-LAPE measurements to the dynamic dielectric response of the sample in the surface region paves the way to monitor electronic and/or lattice dynamics in this region using the extreme site-specificity provided by XPS.

\section{\label{sec:Experimental}Experimental }

The tr-XPS experiment was carried out at the plane-grating monochromator beamline (PG2) at FLASH \cite{martins_2006,gerasimova_2011}, using the photoemission end-station WESPE (Wide-angle Electron SPEctrometer). A SPECS Themis 1000 time-of-flight analyzer equipped with a segmented, position-sensitive delay-line detector (DLD8080-4Q, Surface Concept) was used to record the photoelectron spectra. FLASH was operated at a fundamental wavelength of 7.5\,nm with an effective repetition rate of 4\,kHz (pulse trains of 400 pulses each with an intra-train repetition rate of 1\,MHz and a train repetition rate of 10\,Hz). The monochromator was operated with the 1200\,lines/mm plane grating, a fix-focus constant (c\textsubscript{ff}) of 2, and was tuned to the third harmonic of the FLASH fundamental in first diffraction order. A photon energy of $\sim\,$514.5\,eV was used, with a bandwidth of 80\,meV. The FEL pulse length on the sample was 140\,fs (FWHM). This value includes the temporal broadening induced by the monochromator grating.
The samples are exposed to 100\,fs (FWHM) long IR laser pulses with a center wavelength of 1030\,nm. They are generated in the PIGLET laser system \cite{Seidel2022} that is operated with the same pulse pattern as FLASH. The laser beam size was, at normal incidence, 220\,$\mu$m x 270\,$\mu$m with an X-ray beam size of about 70\,$\mu$m x 70\,$\mu$m, ensuring good spatial overlap and homogeneous laser intensities across the probed sample area. The optical laser fluence on target was $\sim\,$12\,$\mu$J/cm\textsuperscript{2}.
The measurement geometry is such, that the FEL and optical laser impinge under p-polarization on the sample at an incidence angle of 55$^{\circ}$ with respect to the surface normal. The spectrometer detects electrons emitted in a cone of $\pm$7$^{\circ}$ under normal emission.

Two different sample materials, platinum (Pt) and tungsten (W), were selected to perform the LAPE experiments . Both 4f transition metals lend themselves to high-resolution sideband measurements - due to the small energy spread of the 4f photoemission lines. Both samples were cleaned by standard procedures: The Pt\,(111) crystal by repetitive cycles of Ar\textsuperscript{+}-sputtering and annealing, the W\,(110) crystal by annealing in an oxygen atmosphere and using high-temperature flashes.

\section{Results and Discussion}

\begin{figure*}[h]
\includegraphics[width=.99\linewidth]{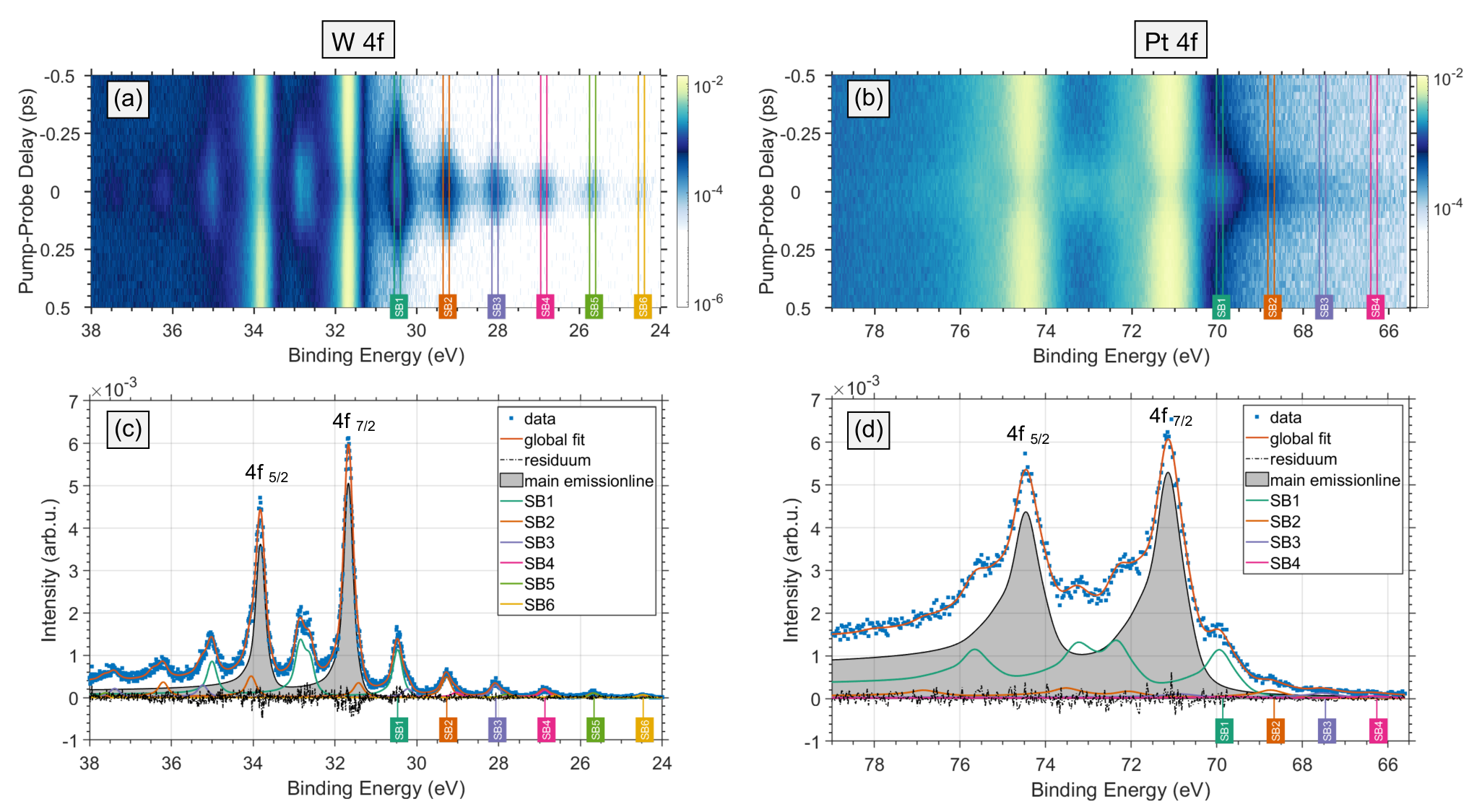}
\caption{Femtosecond time-resolved XPS spectra of the (a) W\,4f and (b) Pt\,4f photolines as a function of time-delay (vertical) and binding energy (horizontal). (c) and (d) Photoelectron spectra of the W\,4f and Pt\,4f core-level obtained in the absence (black line - gray shaded area) and presence (blue marker) of the IR laser pulse. The red line represents a convolution of the unperturbed photoelectron spectrum with the response function (RF). The black dashed line indicates the difference between the fit and the data.}
\label{fig1}
\end{figure*}

Figure\,\ref{fig1} (a) shows femtosecond time-resolved X-ray photoemission spectra of W 4f photolines as a function of time-delay (vertical) and binding energy (horizontal). The pump-probe delay range is $\pm$\,0.5\,ps, whereby negative values are associated with the FEL pulse arriving before the optical laser pulse. The two main features at 31.6\,eV (4f\textsubscript{7/2}) and 33.8\,eV (4f\textsubscript{5/2}) correspond to the two spin-orbit components of the W 4f core-level. 
The unperturbed photoelectron spectrum is significantly modified under the presence of the IR laser field. This is depicted in Fig.\,\ref{fig1}\,(c) by a lineout extracted from Fig.\,\ref{fig1}\,(a) at a pump-probe delay of 0ps. Distinct sidebands up to the sixth order appear on both sides of the central main peak, separated from it by integer mutiples of the IR photon energy (1.2\,eV). 
For the 4f\textsubscript{7/2} line, the energy positions of the sidebands are indicated by boxes with different colors (cf. SB\,1 - SB\,6 in Fig.\,\ref{fig1} (a) and (c)).

For comparison, Fig.\,\ref{fig1}\,(b) and (d) show tr-XPS spectra for Pt\,(111) recorded under the same experimental conditions. Again, the full pump-probe dataset (see Fig.\,\ref{fig1}\,(b)) is dominated by the two spin-orbit components of the 4f core-level located at 71.2\,eV (4f\textsubscript{7/2}) and 74.5\,eV (4f\textsubscript{5/2}). Similar to the tungsten measurement, sidebands are observed in the region of temporal overlap between the optical pump and the X-ray probe pulse (see Fig.\,\ref{fig1}\,(d)). However, a closer inspection of the results for W\,(110) and Pt\,(111) reveals significant differences. 

First, the main photolines and sidebands have different widths in the energy domain, whereby the W 4f lines are significantly narrower compared to the Pt 4f lines. This difference can be partly explained by the different life-time widths of the 4f vacancies in W and Pt. Indeed, according to XPS measurements, the width of the 4f$_{7/2}$ vacancy in Pt is about 5 times larger than in W (see, for example \cite{Campbell01} and references therein). We note that for both materials surface core-levels could not be distinguished because of the reduced surface sensitivity at the chosen excitation energy of 514.5\,eV.

Second, the number of sidebands observed under identical experimental conditions differs significantly for the two materials. For tungsten, we observe sidebands up to the sixth order, whereby for platinum the maximum observe sideband order is four. This observation is somehow surprising and has never been reported before. We note that the kinetic energy of the 4f photoelectrons in both cases is comparable (439\,eV for Pt and 479\,eV for W), and it is much greater than the energy of the IR photons. Below, we provide a semi-quantitative explanation of such a difference and show that the difference in the number of sidebands is related to the different optical properties of solid materials.
 
To quantify the observed LAPE effects, we use a phenomenological description similar to the one used in \cite{miaja-avila_2006}. In a first step the unperturbed emission line was fitted. Here, a dataset well before the temporal overlap of optical and FEL pulses, between -0.9\,ps and -0.7\,ps delay, was selected to ensure the reference spectra are not impacted by the optical laser. Especially the optical laser induced space charge could play a significant role, as it influences the emission spectrum by an energy shift and peak broadening already well before the true temporal overlap \cite{oloff_2014}. The fit was done simultaneously for both spin-orbit components with a Doniach-Sunjic shape \cite{Doniach1969} representing the natural line-shape of the tungsten 4f core-levels. This line shape was convoluted with a Gaussian function to take into account the energy broadening induced by the spectrometer and the X-ray photon energy spread after the monochromator.

Reminiscent of the approach in \cite{miaja-avila_2006}, we convoluted the unperturbed photoemission spectrum with a response function, which consists of a series of Gaussian shaped functions spaced by the laser photon energy $\hbar\omega_{IR}$. The response function reads:

\begin{equation}\label{eq:response_func}
RF(E) = \left( \frac{1- \sum^n_{j=1,\pm}A_{j,\pm}}{\sqrt{2\pi \sigma^2}} \right) e^{-(E-E_0)^2/2\sigma^2} + \sum^n_{j=1,\pm}\frac{A_{j,\pm}}{\sqrt{2\pi \sigma^2}}e^{-(E-E_{0,SB}\pm j\hbar\omega_{IR})^2/2\sigma^2}.
\end{equation}

Here, the first term replicates the unperturbed spectrum at energy $E_0$ with the intensity reduced by the fraction of electrons that populate the sidebands. The second term represents the sidebands, which are described by scaled copies of the central peak, shifted in energy by integer multiples of the IR photon energy. The Gaussian functions used to describe the sideband peaks are separated in energy by integer values of the optical photon energy and they are broadened with a width $\sigma$ that reflects laser related specifics such as the spectral bandwidth and space charge effects. All scaling factors $A_{j,\pm}$ are determined with the overall photoemission intensity remaining constant. A fit with the above response function is performed for the W and Pt spectra.

\begin{figure*}[t]
\includegraphics[width=.99\linewidth]{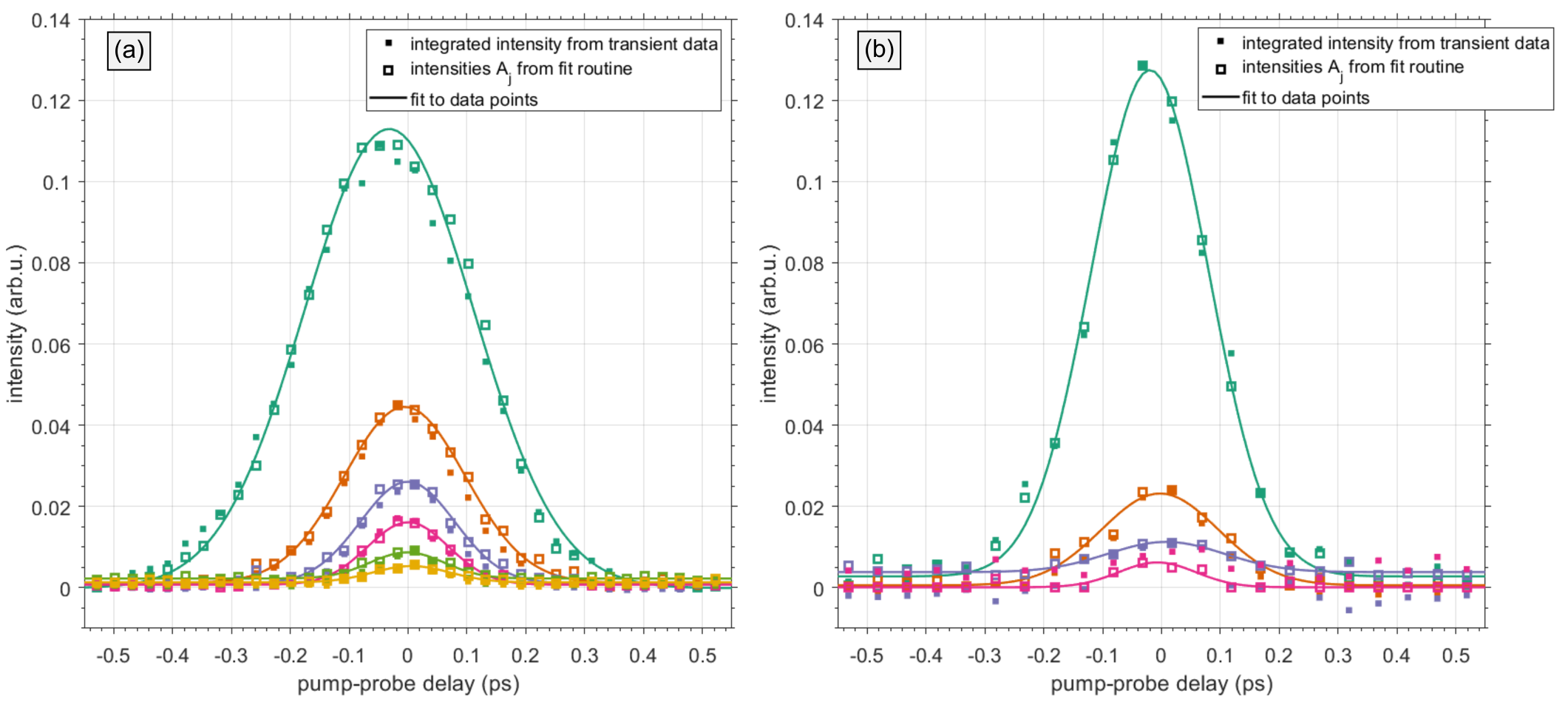}
\caption{Lineouts of the transient signal at constant energy intervals associated with SBs for W 4f (a) and Pt 4f (b) generated at exactly same conditions. The solid marker represent the intensity from integration within fixed energy regions (cf. marked areas of the different sidebands as shown in Fig.\,\ref{fig1}). The open marker display the resulting SB intensities $A_j$ of the fit routine described in detail in the text. Solid lines are fits of the $A_j$ temporal evolutions to Gaussian functions. The selected color code is identical to that in Fig.\,\ref{fig1} and corresponds to the different observed SB orders.}
\label{fig2}
\end{figure*}

Red lines in Fig.\,\ref{fig1} (c) and (d) show the results of the fit as well as the components for all observed sidebands for W 4f and Pt 4f, respectively. Moreover, Fig.\,\ref{fig2} (a) and (b) show lineouts of the transient signal, extracted from the SB regions marked in Fig.\,\ref{fig1} (a) and (b), in comparison to the intensities $A_j$. The integrated transient data are adjusted in intensity to match the corresponding $A_j$. Solid lines in Fig.\,\ref{fig2}\,(a) and (b) are the results of fitting the temporal evolution of the $A_j$ parameters with Gaussian functions. The results in Fig.\,\ref{fig2} show that the parametrization of the spectra using the RF function according to eq.\,(1) provides an excellent description of the SB spectra for all SB orders and all pump-probe delays.\\

In order to elucidate the physical origins of the observed trends, we use a theoretical model of short-pulse two-color photoemission based on the Strong Field Approximation \cite {Keldysh65}. Note that the penetration depths of both light pulses are much larger than the electron inelastic mean free path near the metal surface. For both targets the latter is 5\,-\,7\,\AA\,\cite{Powell99}. Therefore only the first few atomic layers contribute to the measured spectra.  The emission of an electron from an atom at the surface is considered as a one-step process within the single-active electron approximation. 

The electron transits from the initial tightly bound 4f state to the final state in continuum distorted by the IR field. The initial state is described as purely atomic 4f state while the final continuum state is represented by a Volkov wave function \cite{Volkov35} which describes the electron moving in the IR field. Details of this approach and the calculations are presented in the Appendix. 

Within the theoretical model, the sidebands are the result of interference of electron waves emitted at the same phase of the IR field, but in different periods. The number of the sidebands can be roughly estimated as $2A_L\sqrt{2E_e}/\omega_L$, where $A_L$ is the vector potential of the IR field and $\omega_L$ is its frequency. $E_e$ is the kinetic energy of the photoelectron \cite{Kazansky10}. Within this model, the number of sidebands is proportional to the value of the laser vector potential, which is considered constant in space and only dependent on time. Since the vector potential is the same and the electron kinetic energy is similar for both targets, one would expect that the number of sidebands is also similar.

 \begin{figure*}[t]
\includegraphics[width=.99\linewidth]{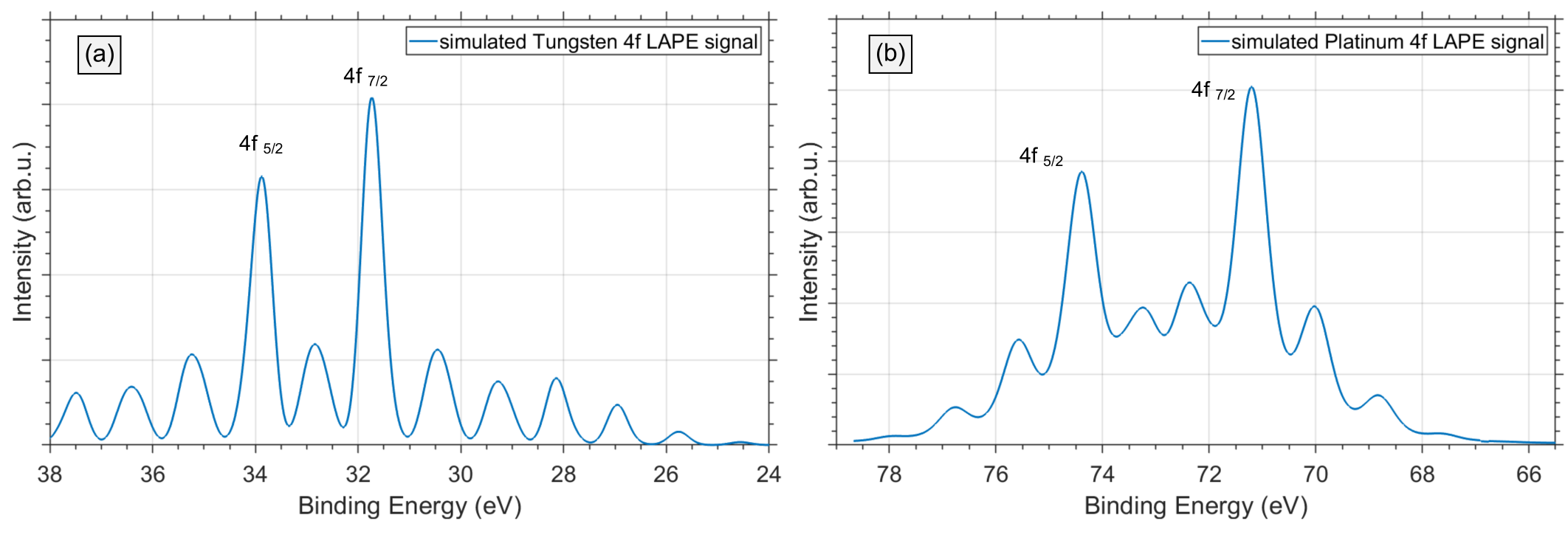}
\caption{Simulated photoemission spectrum for W 4f photolines (a) and Pt 4f (b). The spectra are shown for the IR peak intensity of $3\times 10^{11}$\,W/cm$^2$ in case of W and $0.7\times 10^{11}$\,W/cm$^2$ in case of Pt.  Details about the simulation can be found in the text and in the Appendix. The energy scale was transfered into binding energy to match the experimental data display.}
\label{fig3}
\end{figure*}

Considering the vector potential as a variable parameter, we calculated the photoelectron spectra from W and Pt with other parameters similar to the experiment (see Appendix). The calculations show that the IR field applied to W would have to be 3-4 times larger than for Pt in order to obtain spectra comparable to those observed in the experiment (see Fig. \ref{fig3}). Such as variation of the IR intensity is clearly not commensurate with the experiment, where the it was kept constant throughout both measurements.

To resolve this discrepancy, we resort to the theory of photoemission from metal surfaces, which was widely discussed toward the end of the last century. It was realized that, in order to describe the experimental photoemission yield, it is necessary to take into account the spatial non-uniformity of the electromagnetic field near the surface \cite{Levinson79}. The $z$ component of the vector potential is discontinuous at the surface and the effects of refraction and reflection must be taken into account. The impact of these effects may be roughly estimated by assuming that the sample acts as an isotropic medium of dielectric constant $\epsilon_m (\omega_L)$, having a sharp surface in the plane $z=0$. In this case, the electric field can be computed using Fresnel equations.

As shown in \cite{Feibelman74,Whitaker78,feibelman_surface_1982}, the $z$-component of the vector potential just below the surface from where the electrons are emitted, may be presented as
\begin{equation}\label{eq:AZ}
A_{Lz} = A_0 \frac{2 \cos\theta_i \sin \theta_i}{\epsilon_m \cos \theta_i +
(\epsilon_m - \sin^2  \theta_i )^{1/2}}
\end{equation}
where $A_0$ is the amplitude of the IR vector potential of the incident
wave, $\theta_i$ is the incidence angle of the IR light and $\epsilon_m$ is the complex dielectric constant of the solid:
\begin{equation}
\epsilon_m = \epsilon_{1m} + i \epsilon_{2m}.
\end{equation}

The dielectric constant strongly depends on the IR photon energy. Its tabulated values can be found, for example, in \cite{Adachi12}. For the IR photon energy of our experiment the dielectric constants are
\begin{eqnarray}
\epsilon_{1m} = - 4.25; \hspace{1cm}\epsilon_{2m} = 21.8 \hspace{1cm} {\rm for \; W} \nonumber\\
\epsilon_{1m} = -22.4; \hspace{1cm} \epsilon_{2m} = 42.0 \hspace{1cm} {\rm for \; Pt} \nonumber.
\end{eqnarray}
Substituting these values into Eq.\,(\ref{eq:AZ}) and taking into account the incidence angle $\theta_i = 55^\circ$ we obtain a ratio of $|A_{Lz}|^2_W/|A_{Lz}|^2_{Pt} =3.77$ for the normal components of the IR vector potential in the different materials. Therefore, the intensity of the IR field just under the surface is almost four times larger in W than in Pt. This finding agrees well with the IR intensity ratio identified above by the parametrization model to explain the different sideband numbers.

The presented results show that the influence of an IR driving field for pump-probe experiments induces complex interaction of several influencing factors. Yet, it is also shown that a careful evaluation of these processes is possible and has to be done in order to compare results from different targets. Additionally, one has to take into account that these screening processes also have a temporal component. The LAPE process is dependent on the actual electric field in the surface region, which differs from the intrinsic field of the laser pulse in terms of both its amplitude and its temporal evolution. This offers new insights not only into the spatial, but also the temporal evolution of electronic screening processes at surfaces.
 
We would like to note that attosecond time-resolved photoemission using the streaking approach has recently been used to gain insight into the propagation and damping of electronic and optical wave packets at solid surfaces like W(110), Mg(0001), or even more complex systems like the Mg/W(110) interface or the van der Waals crystal WSe$_2$ \cite{Cavalieri_2007,Neppl_2015,Neppl_2012,Siek_2017}. In all cases, time delays between photoelectrons from different emission channels were recorded and several theoretical studies were published to explain those observations. However, the detailed dielectric response of the sample to the IR laser field was not taken into account in these studies.

The demonstrated sensitivity of the presented femtosecond tr-LAPE measurements to the dielectric properties of the sample in a sub-nm thick surface layer could be exploited to monitor electronic and lattice dynamics in this surface region and provide a deeper understanding of the theory of surface electromagnetic fields including local fields at a chemisorbed atom or molecule. Moreover, using higher X-ray photon energies—larger electron escape depths—one may also be able to probe properties and dynamics at buried interfaces as long as the IR field still reaches these sample regions. 

The findings presented here demonstrate that the presence of the laser-assisted photoelectric effect in high-resolution pump-probe core level spectroscopy requires special attention to disentangle the specific contributions of resonant and non-resonant effects. Yet, the results also provide new opportunities for even more challenging experimental and theoretical investigations utilizing ultrafast and intense laser fields at next-generation high repetition-rate X-ray light sources.

\section{Summary}
In summary, we present a systematic investigation of the laser-assisted photoelectric effect in two similar metallic solids - W\,(110) and Pt\,(111) single crystals. The unexpected difference in the numbers of sidebands generated under the same optical laser and X-ray source conditions can be semi-quantitatively y explained by the sensitivity to screening effects and the dynamic dielectric response of solid state materials. Using the strong-field approximation, we show that the intrinsic, near-surface IR laser field enhancement by a factor of 4 is the root cause for the larger number of sidebands in tungsten compared to platinum.

\section*{Data availability}

The datasets that support the findings of this study are available from the corresponding authors on reasonable request.

\section*{Author contributions}

The experiment was conceived by L.W. and F.R., in discussion with W.E., F.P., D.K., and G.B.. L.W., S.P., D.K., and F.R. planned the experiment and prepared the samples. The experiments were performed by L.W., S.P., D.K., F.P., M.S., D.P., N.W., M.B., and F.R. with input from W.E., O.G., S.D., and S.M. The data analysis and physical interpretation was performed by L.W., N.M.K., F.R., S.D., and W.E. All authors participated in discussing the data. The paper was written by F.R., L.W., and N.M.K. with input from all authors.

\begin{acknowledgments}
We acknowledge DESY (Hamburg, Germany), a member of the Helmholtz Association HGF, for the provision of experimental facilities. Beamtime was allocated with the Application ID 11010397. Moreover, we would like to thank the staff at FLASH at DESY for their excellent support during the experiment. We thank H. Meyer and S. Gieschen from University of Hamburg for support with the instrumentation. These experiments were supported within the research program 'structure of matter' of the Helmholtz Gemeinschaft (HGF). F.R. acknowledges financial support from DESY. This work was supported by the BMBF (Grant No. 05K22OF2  within ErUM-Pro). N.M.K. acknowledges the hospitality and financial support from Donistia International Physics Center (DIPC) as well as financial support from European XFEL. He is also grateful to E.E. Krasovskii and V.M. Silkin for many illuminating discussions. N.W. acknowledge funding by the DFG within the framework of the Collaborative Research Centre SFB 925 - 170620586 (project B2). M.B. and O.G. were supported by the Atomic, Molecular, and Optical Sciences Program of the U.S. Department of Energy, Office of Science, Office of Basic Energy Sciences, Chemical Sciences, Geosciences and Biosciences Division, through Contract No. DE-AC02-05CH11231.
\end{acknowledgments}

\section*{Appendix}

\subsection{Strong field approximation}

We consider photoionization of an atom at a metallic surface in the field of two pulses:
a comparatively weak XUV pulse and a rather strong IR pulse. The pulses co-propagate along the same direction at angle $\theta_i$ (measured from a surface normal) and are linearly polarized in a plane of incidence ("p-polarized").
For the description of ionization of the atoms by a combined action of the XUV and the IR fields, we use a well known approach based on the Strong Field Approximation (SFA) \cite{Keldysh65}. The emission of an electron is considered as a one-step process of transition from the initial bound state to the final 
continuum state distorted by the IR field. The interaction of the electron with the XUV pulse is described within the first-order time-dependent perturbation approach with usage of the Rotating Wave Approximation (RWA). Within the single-active-electron approximation the ionization amplitude can be presented as \cite{Kazansky10} (atomic units are used throughout unless otherwise indicated)
\begin{eqnarray}\label{eq:amplitude1}
{\cal A} (\vec k) \sim -i \int_{-\infty}^{\infty} dt \; \tilde {\cal
E}_X(t) <\psi_{\vec k}(t)|\hat d|\psi_0> e^{i(|E_b| -
\omega_X)t}\,,
\end{eqnarray}
where $\tilde {\cal E}_X(t)$ is the envelope of the XUV pulse,
$\omega_X$ is its mean frequency, $E_b$ is the binding
energy of the electron, $\hat d$ is the dipole operator,
which describes the interaction of the
electron with the XUV field. The function $\psi_0$ is the
initial single-electron wave function. Since in the experiment ionization
of an inner atomic shell is studied, we assume that $\psi_0$ is a localized
function of the atomic electron. Thus we ignore a solid-state environment.

The wave function $\psi_{\vec k}(t)$ in equation
(\ref{eq:amplitude1}) describes the "dressed" photoelectron in the
IR field, which is characterized by the final (asymptotic)
momentum $\vec k$. It is represented by the non-relativistic Volkov wave
function \cite{Volkov35}:
\begin{equation}\label{eq:Volkovfunc}
\psi_{\vec k} = \exp \left\{i[\vec k-\vec A_L(t)] \vec r -i\Phi_V(\vec
k,t)\right \}\,.
\end{equation}
Here $\Phi_V(\vec k,t)$ is the Volkov  phase
\begin{equation}\label{eq:Volkovphase}
\Phi_V(\vec k,t) = -\frac{1}{2} \int^{\infty}_t dt' \left[ \vec k -
\vec A_L(t') \right ]^2 \,
\end{equation}
with $\vec A_L(t)$ being the vector potential of the IR field,
which is defined as $\vec A_L(t)= \int_t^\infty dt'\vec {\cal
E}_L(t')$, where $\vec {\cal E}_L(t)$ is the IR electric field
vector. As a first approximation we assume that the IR vector potential
is spacial constant.

\subsection{Evaluation of the dipole matrix element}

In the considered case, the initial electron state
is $4f$-state with orbital angular moment $l_0=3$.
The final continuum sate $\psi_{\vec k_0}$, where $\vec k_0 = \vec k -A_L(t)$ may be expanded in partial waves.
The dipole transition is possible to the "g" and "d" partial waves.
Since at high photoelectron energy the transition to the "$l_0+1$" state
dominates, we ignore the transitions to the $d$-continuum. Then the
matrix element in equation (\ref{eq:amplitude1}) for a particular projection of
the initial orbital angular momentum $m_0$ may be presented as
\begin{eqnarray}\label{eq:matrixel}
<\psi_{\vec k_0}(t)|\hat d|\psi_{4f,m_0}> \sim e^{i\Phi_V(\vec k,t)}
Y_{4,m_0}(\theta_0,\phi_0) {\cal R}^1_{k_0,l_0} C (l,l_0,m_0) \,,
\end{eqnarray}
where $Y_{l,m_0}(\theta_0,\phi_0)$ is the spherical harmonic, ${\cal R}^1_{k_0,l_0}$ and $C(l,l_0,m_0)$ are the radial and angular parts
of the matrix element. $C(4,3,m_0) = <Y_{4,m_0}|Y_{1,0}|Y_{3,m_0}>$. 
The angles $(\theta_0, \phi_0)$ give
the direction of electron emission from the atom before propagation
in the optical laser field. These angles are connected with the
detection angles $(\theta, \phi)$ after propagation in the IR field
by the relations:
\begin{eqnarray}\label{eq:angles}
\theta_0(t) = \arcsin (k \sin \theta/k_0(t))\,, \\
k_0^2(t) = (\vec k - \vec A_L(t))^2 \,.
\end{eqnarray}

Substituting the matrix element into the amplitude (\ref{eq:amplitude1})
one gets
\begin{eqnarray}\label{eq:amplitude2}
{\cal A}_{m_0} (\vec k) \sim -i C_{l,l_0,m_0}
 \int_{-\infty}^{\infty} dt \; \tilde {\cal
E}_X(t) e^{i\Phi_V(\vec k,t)} {\cal R}^1_{k_0,l_0}
Y_{4,m_0}(\hat{k}_0) e^{i(|E_b| -
\omega_X)t}\,,
\end{eqnarray}
which gives for the cross section
\begin{equation}
\frac{d\sigma}{d\Omega}  (\vec k) = \sum_{m_0} |{\cal A}_{m_0} (\vec k) |^2 \,.
\end{equation}

\subsection{Details of calculations}

The envelope of the XUV pulse (electric field) is Gaussian:
\begin{equation}\label{eq:xuv}
  \tilde{\cal E}_X =  {\cal E}_0 e^{-(t-T_{init})^2/T_{XUV}^2}
\end{equation}
where $T_{XUV} = 3018 $ a.u. (duration of the XUV pulse 
FWHM $\sim 80$ fs)
and $T_{init} = 970 $ a.u. (the middle of the IR pulse)

The envelope of the IR pulse (electric field):
\begin{eqnarray}\label{eq:ir}
  \tilde {\cal E}_L=   {\cal E}_{L0}/(1 + e^{(-t/T_0)}) \hspace{2cm} {\rm if} \; t \le T_{IR}/ 2\\
          =  {\cal E}_{L0}/(1 + e^{ ((t-T_{IR})/T_0)} )\hspace{2cm} {\rm if} \; t \ge  T_{IR}/2
\end{eqnarray}
where $ {\cal E}_{L0}$ is a peak value of the IR electric field,
$T_0 = 300$ a.u. , $T_{IR} = 1940 $ a.u (duration of the IR pulse FWHM $\sim 50$ fs). The IR frequency is 0.0441 a.u. (1030\,nm).
Note, that by calculation reasons, the duration of both pulses is
about two times smaller than in the experiment.

At the photon energy 515\,eV, the kinetic energy of an electron emitted from 4f subshell of W is taken as 17.5\,a.u. (476\,eV).
No fine-structure splitting is included.
For Pt the kinetic energy  is 16.17\,a.u. (440\,eV).

The calculations have been done for an emission angle (with respect to the photon electric field) $35^\circ$ which corresponds to the normal emission, perpendicular to the surface. Since at high electron energy the radial parts of the dipole matrix elements are slowly varying function of energy, in numerical calculations we consider them as a constant and put them to unity. 

The calculated spectra were additionally convoluted with the Gaussian, imitating the energy resolution of the detector and the spectral width of the laser pulse
which were assumed to be about 0.4\,eV. Additionally, the spectra are convoluted with Lorentian with $\Gamma = 0.06$ eV for W and 0.4\,eV for Pt. 
To imitate the spin-orbit splitting we added to the calculated spectrum its copy shifted in energy by 2.1\,eV for W and 3.2\,eV for Pt and diminished by factor of 0.75. The resulting spectra are shown in Fig\,\ref{fig3}.

\bibliography{LAPE_Ref}
\bibliographystyle{apsrev4-1}


\end{document}